\documentclass[twocolumn,amsmath,amssymb,nofootinbib,superscriptaddress]{revtex4}
\usepackage{amssymb}
\usepackage{graphicx}
\usepackage{longtable}
\usepackage{amsmath}
\usepackage{amsfonts}
\usepackage{color}
\usepackage{epsfig}
\usepackage[latin1]{inputenc}

\begin{document}

\date{\today }

\title{Manipulation of spin cluster qubits by electric-field induced modulation \\ of exchange coupling, $g$-factor, and axial anisotropy}

\author{Filippo Troiani}
\email{filippo.troiani@nano.cnr.it}
\affiliation{Centro S3, CNR-Istituto di Nanoscienze, 
I-41125 Modena, Italy}

\begin{abstract} 
Electric fields are increasingly used for coherently manipulating spin states in semiconductor and molecular systems. Here we discuss the spin manipulation allowed by the modulation of the main parameters entering the Hamiltonians of molecular spin clusters. In particular, we focus on transitions between states that differ in terms of scalar quantities, such as the total- or the partial-spin sums, but have vanishing expectation values of both the total- and the single-spin projections. These conditions supposedly define subspaces that are immune to the main decoherence mechanisms and that cannot be identified in individual spins, where the total- and the partial-spin sums are fixed or cannot be defined. We show that the desired transitions can in principle be realized in systems as simple as a spin dimer, by modulating the $g$-factor, the axial anisotropy, or, under suitable conditions, the exchange interaction.
\end{abstract}

\maketitle

\section{Introduction}

The standard approach for coherently manipulating electron spins is based on the use of oscillating magnetic fields \cite{Schweiger2001a}. However, the coherent control of individual spins requires extreme spatial and temporal resolutions, which make electric fields preferable. Electron and hole spins in semiconductor systems can be efficiently manipulated by oscillating electric fields in the presence of spin-orbit coupling, either intrinsic or induced by different means \cite{Kato2003a,Nowack2007a,Mi2018a,Crippa2018a,Pingenot2008a,Kato2003a,Takahashi2013a,Schroer2011a}. 

Purely spin systems, such as paramagnetic defects and molecular nanomagnets \cite{Gatteschi2006a,Furrer2013a}, are characterized by approximately frozen orbital degrees of freedom. In spite of this, spin-electric coupling and coherent control of the spin state by electrical means have been demonstrated in the last years \cite{George2013a,Godfrin2017a,Boudalis2018a,Fittipaldi2019a,Liu2019a}. From a theoretical point of view, the effect of an applied electric field in a molecular nanomagnet can be described by the modulation of the parameters that enter the spin Hamiltonian. In the case of individual spins, such modulation typically induces transitions -- or renormalizes the energy gap -- between Zeeman levels \cite{Kato2003a,Pingenot2008a,Schroer2011a,George2013a,Takahashi2013a,Crippa2018a}. Spin clusters provide additional possibilities for the spin-electric coupling and for the resulting coherent control of the spin state. Electric fields can in fact modulate the exchange constants \cite{Fittipaldi2019a}, thus enabling the spin manipulation even in the absence of a sizeable spin-orbit interaction \cite{Trif2008a}. Besides, through spin-electric coupling in spin clusters one can generate linear superpositions between eigenstates that differ in terms of scalar quantities and share the same value of the total spin projection. This supposedly makes them immune to the main decoherence mechanisms \cite{Troiani2012a,Escalera2018a} and represents one of the potential advantages of spin cluster qubits \cite{Meier2003a}.

Here we consider transitions induced by the modulation of the main terms in typical spin Hamiltonians, namely those accounting for exchange interaction, axial anisotropy and Zeeman coupling. The modulation of the exchange interaction allows the manipulation of the spin chirality in spin clusters formed by an odd number of half integer spins with antisymmetric exchange interactions \cite{Trif2010a}. Here we show that such modulation can be exploited for inducing analogous transitions in a wider class of systems, including simple spin dimers. The modulation of the axial anisotropy and of the Zeeman coupling also induces transitions between states that differ in terms of the total spin, within subspaces of given spin projection. Overall, the spin-electric coupling offers a versatile route for the coherent manipulation of the quantum state in spin clusters, where neither oscillating nor static magnetic fields are in principle required. Besides its fundamental interest, this potentially represents an important motivation for the implementation of molecular spin cluster qubits.

The paper is organized as follows. In Section II we define the approach and the kind of spin manipulation we are interested in. Sections III, IV, and V are focused on the spin manipulation allowed by the modulation respectively of the $g$-factor, the exchange coupling, and the axial anisotropy. Section VI is devoted to the spin manipulation in spin dimers and to the possible role played by the mixing of different $S$ multiplets in the eigenstates. Finally, Section VII contains the conclusions and a brief discussion on the specific differences between this approach and related ones that are presented in the literature.

\section{General setting}

We write the Hamiltonian of the spin cluster as the sum of four different terms:
\begin{equation}
H = H_a + H_b + \delta H_a (t) + \delta H_b (t) , 
\end{equation}
where $H_a$ commutes with both ${\bf S}^2$ and $S_z$ and typically represents the dominant part of the system Hamiltonian, while $H_b$ doesn't commute with ${\bf S}^2$ and/or $S_z$. Finally, the terms $\delta H_a$ and $\delta H_b$ account for the changes introduced respectively in $H_a$ and $H_b$ by the application of a time-dependent electric field. 

To be more specific, we consider the case where the dominant terms are given by symmetric exchange interactions between the $N$ spins that form the cluster, and by their homogeneous coupling to an external magnetic field ${\bf B} = B \hat{\bf z}$:
\begin{equation}
H_a = \frac{1}{2} \sum_{k \neq l} J_{kl} {\bf s}_k \cdot {\bf s}_l + \mu_B B \sum_{k=1}^N g_k s_{k,z} , 
\end{equation}
with $g$-factors that are all identical ($g \equiv g_k$ for all $k$) and $J_{lk}=J_{kl}$. The eigenstates of $H_a$ are thus also eigenstates of ${\bf S}^2$ and $S_z$, and can be written in the form $|\alpha,S,M\rangle$. Here, the index $\alpha$ orders the eigenstates within the subspace $(S,M)$ and/or accounts for additional quantum numbers, such as the partial spin sum \cite{Gatteschi2006a}. 

The low-symmetry Hamiltonian $H_b$ can in principle include different kinds of terms \cite{Furrer2013a}. Here, we explicitly mention the single-spin axial anisotropies, which can play an important role in the spin manipulation: 
\begin{equation}
H_b = \sum_{k=1}^N d_k s_{k,z}^2 .
\end{equation}
In the following, unless differently specified, we assume that $H_b$ is small compared with $H_a$, and that the system eigenstates take the above mentioned form $|\alpha,S,M\rangle$.

The expressions of $\delta H_a$ and $\delta H_b$ are obtained from those of $H_a$ and $H_b$, by replacing the coefficients $J_{kl}$, $g_k$, and $d_k$ with the respective, electric-field induced variations $\delta J_{kl}$, $\delta g_k$, and $\delta d_k$. We consider a time-dependent electric field of the form 
\begin{equation}
E(t) = \mathcal{E} f(t) \cos (2\pi \nu_E t), 
\end{equation}
where $\mathcal{E}$ is the maximal amplitude of the field, $0\le f(t)\le 1$ defines its modulation in time, and $\nu_E$ is the frequency at which the field oscillates. The renormalization $\delta x$ of the generic parameter $x$ entering the spin Hamiltonian is assumed to depend linearly on the electric field, and to carry the same kind of time dependence: 
\begin{equation}
\delta x (t) = \delta_{x} f(t) \cos (2\pi \nu_E t). 
\end{equation}
Physically, this corresponds to the assumption that different spin states carry slightly different electric dipoles, such that the first-order contribution is finite and represents the dominant term in the renormalization of the coupling constants.

The modulation of the parameters allows both a resonant and a non-resonant manipulation of the spin cluster qubit. In the former case, the electric field induces transitions between the eigenstates $|0\rangle = |\alpha_0, S_0, M_0\rangle$ and $|1\rangle = |\alpha_1, S_1, M_1\rangle$ of $H_a$, with energies $E_0$ and $E_1$, provided that 
\begin{equation}
\langle 0|\delta H_a + \delta H_b |1 \rangle\neq 0, \ \text{with}\ \nu_E = |E_1-E_0|/h. 
\end{equation}
In the latter case, the electric field allows one to renormalize the gap between the two states, and thus to modify in a controlled fashion the phase accumulated between the two during a given waiting time. The required condition now reads:
\begin{equation}\label{nonres}
\delta E_0 \neq \delta E_1 ,\ \text{with}\ \nu_E=0,
\end{equation} 
where
$\delta E_p = \langle p |\delta H_a + \delta H_b| p \rangle$, with $p = 0,1$. In general, $|0\rangle$ can be identified with the ground state of the spin cluster, whereas $|1\rangle$ corresponds to the lowest excited state that is accessible from $|0\rangle$, according to the selection rules. 

One of the potential advantages offered by spin clusters with respect to individual spins is represented by the possibility of manipulating the system state within subspaces that are relatively immune to decoherence. In fact, transitions induced by electron spin resonance allow the generation of linear superpositions between eigenstates with different values of $M$ \cite{Schweiger2001a}. However, neighboring electron and nuclear spins randomize and efficiently ``measure'' the value of the spin projection, and thus tend to destroy the phase coherence between two eigenstates $|0\rangle$ and $|1\rangle$ with $M_0 \neq M_1$ \cite{Suter2016a}. For this reason, we focus hereafter on linear superpositions between states with $M \equiv M_0 = M_1$. In other words, we consider the manipulation of the quantum numbers $\alpha$ or $S$, within a subspace of given $M$.  

If the environment couples to the spins collectively through the total spin projection, the states with a given $M$ define a decoherence-free subspace \cite{Zanardi1997a,Lidar1998a}. In molecular spin clusters, however, the coupling of the electron spins with the nuclear environment has a rather local character. The above argument should thus be applied not only to the total spin projection, but also to the distribution of the magnetization within the spin cluster \cite{Troiani2008a,Troiani2012a}. The initial and final states of the electric-field induced transition are thus required to be indistinguishable also in terms of single-spin magnetization:
\begin{equation}\label{eq01}
\langle 0 | s_{k,z} | 0 \rangle = \langle 1 | s_{k,z} | 1 \rangle \ \ (k=1,\dots,N).
\end{equation}
(Being $|0\rangle$ and $|1\rangle$ eigenstates of $S_z$, an analogous condition is automatically fulfilled by the expectation values of $s_{k,x}$ and $s_{k,y}$, which vanish identically.) We note that such condition marks a difference with respect to systems such as the ones that exhibit a toroidal momentum \cite{Chibotaru2008a,Guo2012a,Tang2006a}, where the ground state expectation value of the total spin vanishes, but that of the individual spin doesn't. While Eq. (\ref{eq01}) can in principle be satisfied by states with arbitrary values of $M$, possible distortions in the spin cluster can in practice result in a dependence of the single-spin magnetization on the system eigenstate. For this reason, we focus hereafter on the case $M=0$, where all the expectation values $\langle s_{k,z} \rangle$ vanish by symmetry.

\section{g-factor modulation and resonance}

The first case we consider is that where the applied electric field renormalizes the $g$-factors of the spins. This results in a time-dependent Hamiltonian 
\begin{equation}
\delta H_{a,g} (t)
= \mu_B B \sum_{k=1}^N \delta g_{k}(t)\, s_{k,z} .
\end{equation}
Modulating the $g$-factor in the presence of a static magnetic field is in principle equivalent to applying a time-dependent magnetic field \cite{Kato2003a,Pingenot2008a,Schroer2011a,Takahashi2013a,Crippa2018a}. In electron spin resonance, one typically induces transitions between the system eigenstates by means of a time-dependent magnetic field that oscillates along a direction perpendicular to the quantization axis \cite{Schweiger2001a}. In such a case, the transitions that are induced in the system are characterized by the selection rules 
$
\Delta S = 0
$
and 
$ 
\Delta M =\pm 1
$. 
The above Hamiltonian $\delta H_{a,g}$ can instead induce transitions between states that have identical values of $M$, with an amplitude given by
\begin{equation}
\langle 0 |\delta H_{a,g} (t)| 1 \rangle 
= \mu_B B \sum_{k=1}^N \delta g_{k}(t)\, 
\langle 0 |s_{k,z}| 1 \rangle .
\end{equation}
The selection rules for such transitions read:
\begin{equation}\label{eq02}
|\Delta S | = 1;
\ 
\Delta M = 0.
\end{equation}
The second selection rule simply results from the fact that $\delta H_{a,g}$ commutes with $S_z$. The first selection rule can be deduced by combining the Wigner-Eckart theorem with symmetry arguments. In fact, from the Wigner-Eckart theorem, it follows that $|\Delta S|\le 1$. From the transformation of the the eigenstates $|\alpha,S,M=0\rangle$ under $\pi$ rotations around any axis in the $xy$ plane it follows that $\Delta S$ must be an odd number (see Appendices A and B for further details). The modulation of the $g$-factors can thus be used to resonantly induce singlet-triplet transitions ($S_0=0$, $S_1=1$) in the spin cluster. 

In order for transitions between different values of $S$ to be possible, the $g$-factor modulations of the $N$ spins are required to be not all identical ({\it i.e.} one should not have that $\delta g_k = \delta g_l$ for all values of $k$ and $l$), otherwise $\delta H_{a,g}$ commutes with $H_a$ and cannot induce transitions between two of its eigenstates. 

On the other hand, from the selection rules given in Eq. (\ref{eq02}) it follows that within the present approach -- unlike in electric-field assisted electron spin resonance \cite{Liu2019a} -- the electric-field induced variation of the $g$-factors cannot be used for the nonresonant manipulation. In fact, in the $M=0$ subspace one has that $\langle \alpha, S, 0 | \delta H_{a,g} | \alpha, S, 0\rangle = 0$ for all values of $\alpha$ and $S$. As a result the renormalizations of the energy levels always vanish 
\begin{equation}
\delta E_p  
= \mu_B B \sum_{k=1}^N \delta g_{k}(t) 
\langle p |s_{k,z}| p \rangle\! =\! 0 \ \ (p\! =\! 0,1),
\end{equation}
and the requirement given in Eq. (\ref{nonres}) cannot be fulfilled. 

\section{Exchange-coupling modulation and resonance}

The electric field can also be used to modulate the exchange couplings between the spins \cite{Baadji2009a,Islam2010a,Nossa2012a}. In particular, a linear effect on the exchange coupling $J_{kl}$ is expected if permanent and non-identical electric dipoles are associated to different values of $({\bf s}_k+{\bf s}_l)^2$. The time-dependent Hamiltonian resulting from the renormalization of the exchange couplings in the spin cluster reads
\begin{equation}
\delta H_{a,J} (t) = \frac{1}{2} \sum_{k \neq l} \delta J_{kl}(t) \, {\bf s}_k \cdot {\bf s}_l . 
\end{equation}
The above Hamiltonian commutes with ${\bf S}^2$ and $S_z$, and thus can be used to manipulate the spin state within a subspace of given $S$ and $M$: 
\begin{equation}\label{eq03}
\Delta S = 0;
\ 
\Delta M = 0.
\end{equation}

On the other hand, $\delta H_{a,J}$ can change the additional $N-2$ degrees of freedom that define the system state, such as the partial spin sums $S_k$ or the spin chirality \cite{Bulaevskii2008a,Wang2019a}, here generically accounted by the index $\alpha$. The case of spin chirality has been extensively discussed in Refs. \cite{Trif2008a,Trif2010a,Troiani2012a}. The partial spin sums ${\bf S}_n = \sum_{k=1}^n {\bf s}_k$, with $n=2,\dots,N-1$ define -- together with $S_N\equiv S$ and $M$ -- a complete basis for the $N$-spin cluster. The exchange operator ${\bf s}_k \cdot {\bf s}_l$ (with $k<l$) can couple two basis states $|S_2,S_3,\dots,S_{N-1},S,M\rangle$ only if 
\begin{equation}
|\Delta S_n | \le 1
\ \text{for}\ k \le n < l, \ 
\Delta S_n = 0 
\ \text{otherwise}.
\end{equation}
If (some of) the sums $S_n$ are good quantum numbers, {\it i.e.} if (some of) the operators ${\bf S}_n^2$ commute with $H_a$, the above equation defines additional selection rules for the transitions induced by the modulation of the exchange operator ${\bf s}_k \cdot {\bf s}_l$. 

The modulation of the exchange coupling can thus be used to resonantly induce transitions between different eigenstates $|0\rangle$ and $|1\rangle$ of $H_a$ (with $S_0=S_1$, $M_0=M_1$, and $\alpha_0 \neq \alpha_1$), with an amplitude
\begin{equation}\label{eq11}
\langle 0 |\delta H_{a,J} (t)| 1 \rangle 
= \frac{1}{2} \sum_{k \neq l} \delta J_{kl}(t) \, \langle 0 | {\bf s}_k \cdot {\bf s}_l | 1 \rangle,
\end{equation}
provided that the relative change of the exchange coupling is not homogeneous. In fact, if $\delta J_{kl}/J_{kl}$ is the same for all values of $k,l=1,\dots,N$, then $\delta H_{a,J}$ commutes with $H_a$, and therefore cannot induce transitions between its eigenstates. 

Concerning the nonresonant manipulation, the selection rules in Eq. (\ref{eq03}) show that the renormalization of the exchange couplings can be used to modify the gap between eigenstates of $H_a$:
\begin{equation}
\delta E_p = \frac{1}{2} \sum_{k \neq l} \delta J_{kl}(t) \, \langle p | {\bf s}_k \cdot {\bf s}_l | p \rangle\ \ (p\! =\! 0,1). 
\end{equation}
This applies also to the case where the relative changes $\delta J_{kl}/J_{kl}$ are all identical. In fact, the nonresonant manipulation doesn't require transitions between the system eigenstates, and can thus be performed also if $\delta H_a$ commutes with $H_a$.

\section{Anisotropy-factor modulation and resonance}

The spin-state manipulation can also exploit the modulation of the axial anisotropy induced by the electric field \cite{George2013a}. This would result in the following time-dependent Hamiltonian:
\begin{equation}
\delta H_{b,d}(t) = \sum_{k=1}^N \,\delta d_k(t)\, s_{k,z}^2 .
\end{equation}
The above Hamiltonian commutes with $S_z$, but not with ${\bf S}^2$, and is thus in principle suitable for inducing transitions between states that belong to the $M=0$ subspace, with amplitude
\begin{equation}
\langle 0 |\delta H_{b,d}(t)| 1 \rangle = \sum_{k=1}^N \,\delta d_k(t)\, \langle 0 | s_{k,z}^2 | 1 \rangle.
\end{equation}
The selection rules that characterize these transitions read:
\begin{equation}\label{eq04}
|\Delta S| = 0,2;
\ 
\Delta M = 0.
\end{equation}
As in the previous cases, the second selection rule simply follows from the commutation of the time-dependent Hamiltonian with $S_z$. The first selection rule can be derived by combining the Wigner-Eckart theorem with  symmetry arguments. In fact, from the Wigner-Eckart theorem, it follows that $|\Delta S|\le 2$. From the transformation of the the eigenstates $|\alpha,S,M=0\rangle$ under $\pi$ rotations around any axis in the $xy$ plane it follows that $\Delta S$ must be an even number (see Appendices A and B for further details). Unlike the cases of the $g$-factor and of the exchange-coupling modulation, here the renormalizations $\delta d_k$ need not be inhomogeneous, because $\delta H_{b,d}$ doesn't commute with $H_a$ even if the parameters $d_k$ are all identical. 

From the above selection rules, it follows that the modulation of the axial anisotropy also allows the nonresonant manipulation, by inducing variations of the energy levels
\begin{equation}
\delta E_p = \sum_{k=1}^N \,\delta d_k(t)\, \langle p | s_{k,z}^2 | p \rangle\ \ (p\! =\! 0,1).
\end{equation}
This can result in a renormalization of the difference between the energy eigenvalues, provided that the eigenstates in question are not both singlet states. In fact, being, by symmetry, $\langle s_{k,z}^2\rangle = s_k(s_k+1)/3$ in all singlet states, from $S_0=S_1=0$ it would follow that $\delta E_0 = \delta E_1$, unlike required by Eq. (\ref{nonres}).

\section{Application to simple spin clusters}

\begin{figure}
\centering
\includegraphics[width=0.85\columnwidth]{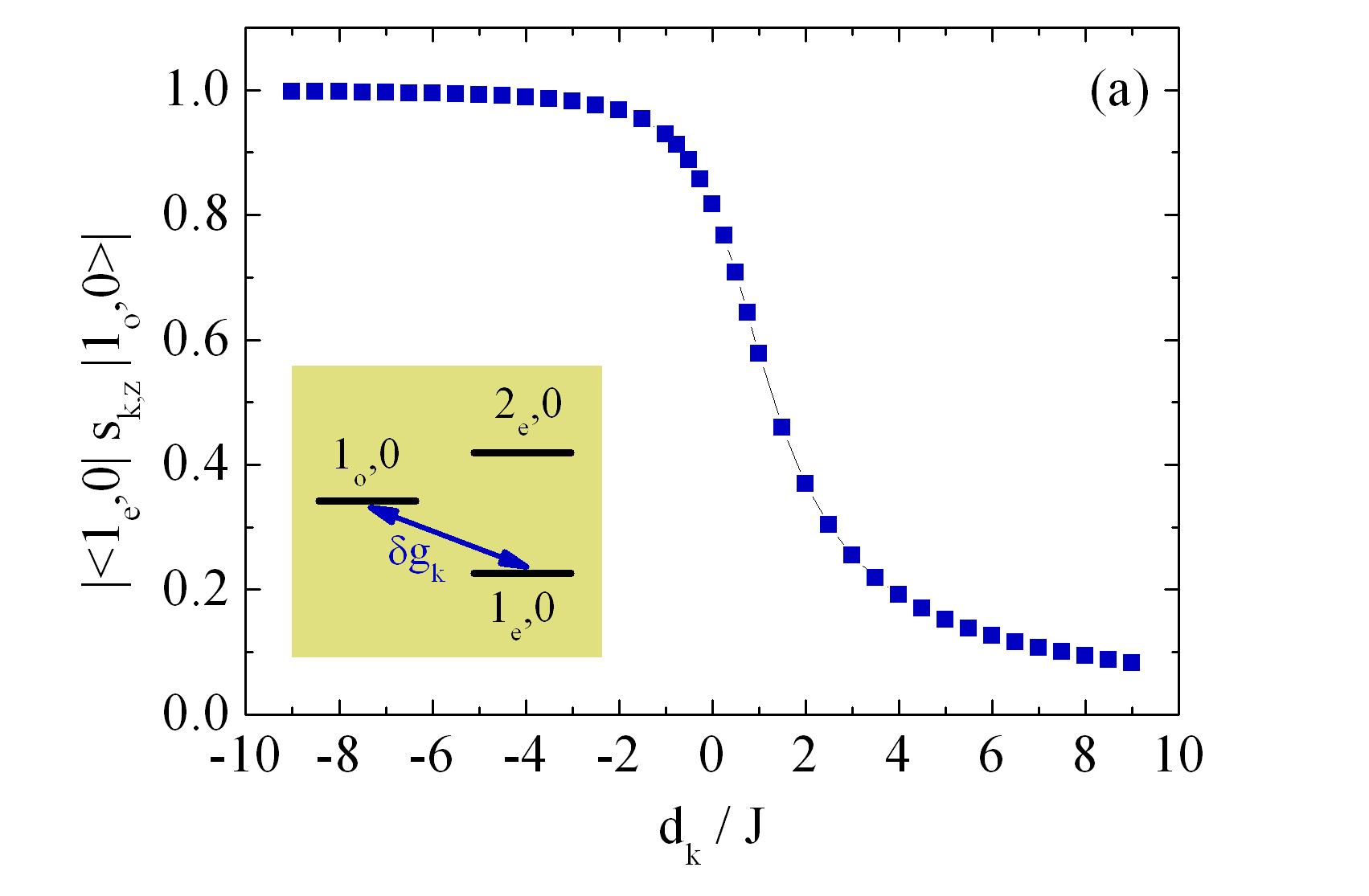}
\includegraphics[width=0.85\columnwidth]{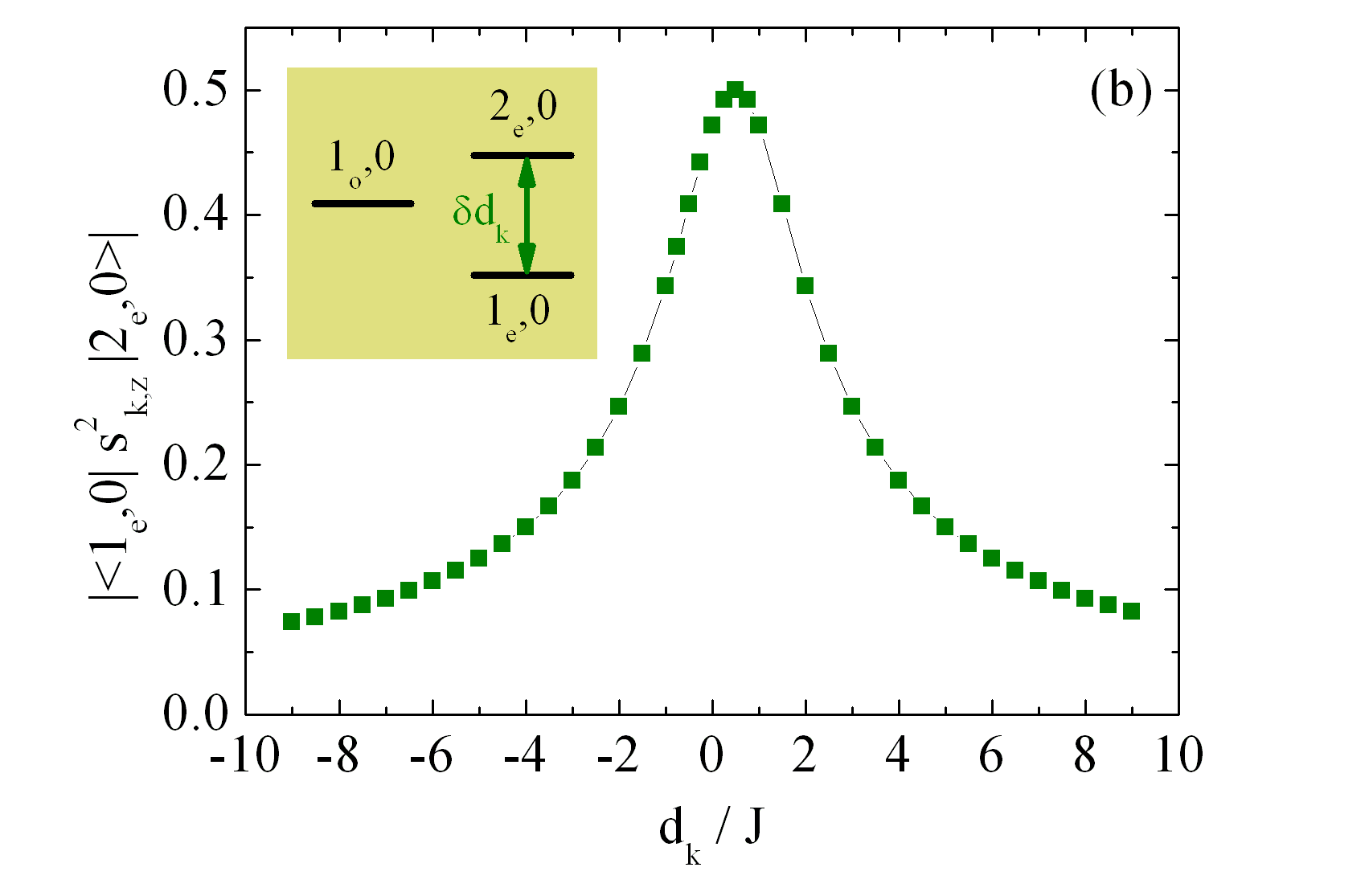}
\includegraphics[width=0.85\columnwidth]{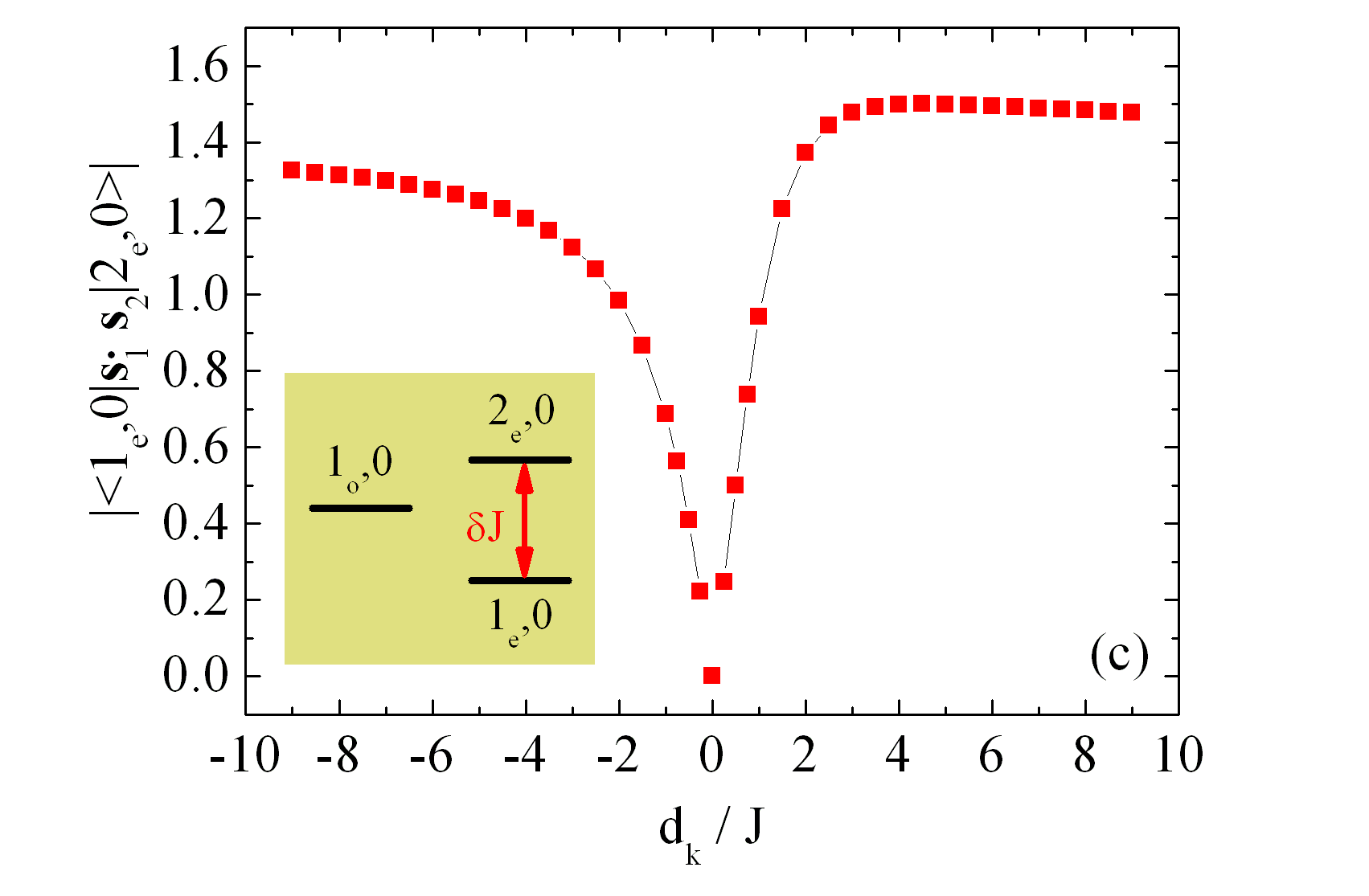}
\caption{\label{fig1} Off-diagonal matrix elements of the operators $s_{k,z}$ (a), $s_{k,z}^2$ (b), and ${\bf s}_1 \cdot {\bf s}_2$ (c) in the spin dimer with $s_1=s_2=1$, as a function of the ratio between the anisotropy ($d$) and exchange ($J$) parameters. These matrix elements determine the amplitudes of the transition between eigenstates $|\beta_\xi,M\rangle$ (with $\xi = e,o$), induced by the modulation of $g$, $d$, or $J$ (figure insets).}
\end{figure}

In the present Section, we consider the simplest possible system where most of the above transitions can in principle take place, namely a spin dimer. We show that also the exchange coupling modulation alone allows resonant spin manipulation in such a dimer, provided that the system eigenstates present a significant degree of mixing between different values of the total spin ($S$-mixing). In the absence of $S$-mixing, the use of exchange-coupling modulation alone is shown to require more complicated spin clusters.

\subsection{Spin dimer qubit in the small-anisotropy limit}

In the case $N=2$ and $s\equiv s_1 = s_2$, most of the transitions considered in the previous Section can in principle take place. The eigenstates of the Hamiltonian $H_a = J {\bf s}_1 \cdot {\bf s}_2 + \mu_B B g S_z$ (with $J>0$) are given by $|S,M\rangle$, with $S=0,\dots 2s$ and no additional index required (being $\alpha \equiv 1$). 

The modulation of the $g$-factor can be used to induce resonant transitions between the ground and first excited states, namely $|0,0\rangle$ and $|1,0\rangle$. The relevant matrix element is that of the single-spin projection, which takes the form (see Appendix C):
\begin{equation}\label{eq09}
\langle 0,0|s_{k,z}|1,0\rangle = (-1)^{k+1} \sqrt{s(s+1)/3} \ \ (k=1,2),
\end{equation}
and thus scales linearly with the length of the constituent spins. As discussed above, in order to make the transition possible, the renormalization of the $g$-factor must be different for the two spins. In a physical system, the inequivalence between the two spins can be introduced, for example, by using a spin heterodimer, where the two spins are provided by two different ions, or the second spin $s_2$ is replaced by two ferromagnetically coupled spins $s_2'$ and $s_2''$, such that $s_2=s_2'+s_2''$.

The modulation of the exchange interaction cannot be used to resonantly induce transitions between the eigenstates in a dimeric system. In fact, in the presence of only one exchange coupling, $\delta H_{a,J}$ always commutes with $H_a$. In order for such transitions to be possible, one needs to consider Hamiltonians where $S$ is not a good quantum number, or spin clusters with $N>2$ (see below). However, $\delta H_{a,J}$ can always renormalize the energy difference between two eigenstates, in order to perform a nonresonant manipulation. In particular, variations in the exchange coupling results in renormalizations of the gaps between $|0,0\rangle$ and $|1,0\rangle$ or $|2,0\rangle$ are by $\delta J$ and $3\delta J$, respectively. 

\begin{figure}
\centering
\includegraphics[width=0.85\columnwidth]{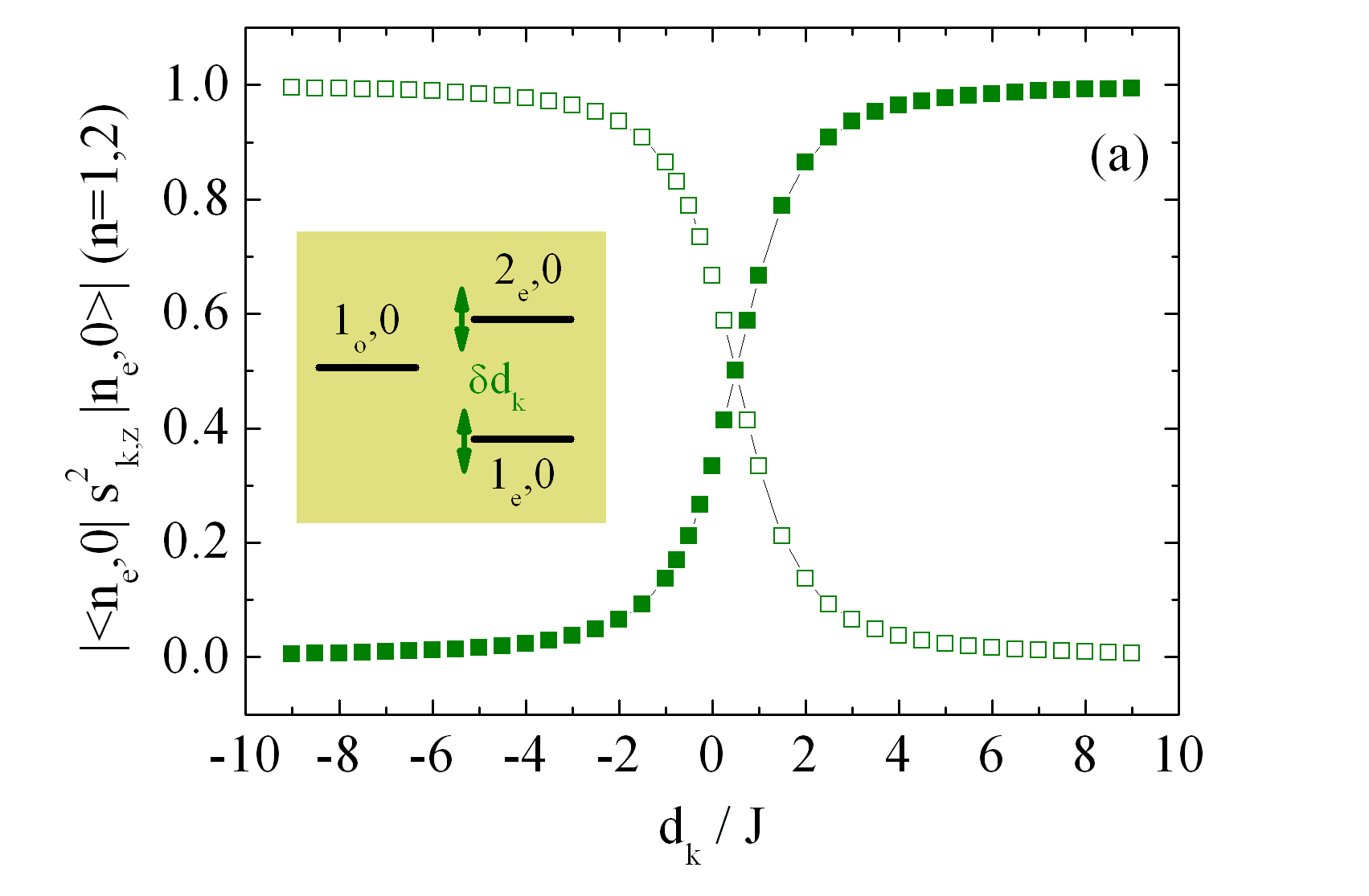}
\includegraphics[width=0.85\columnwidth]{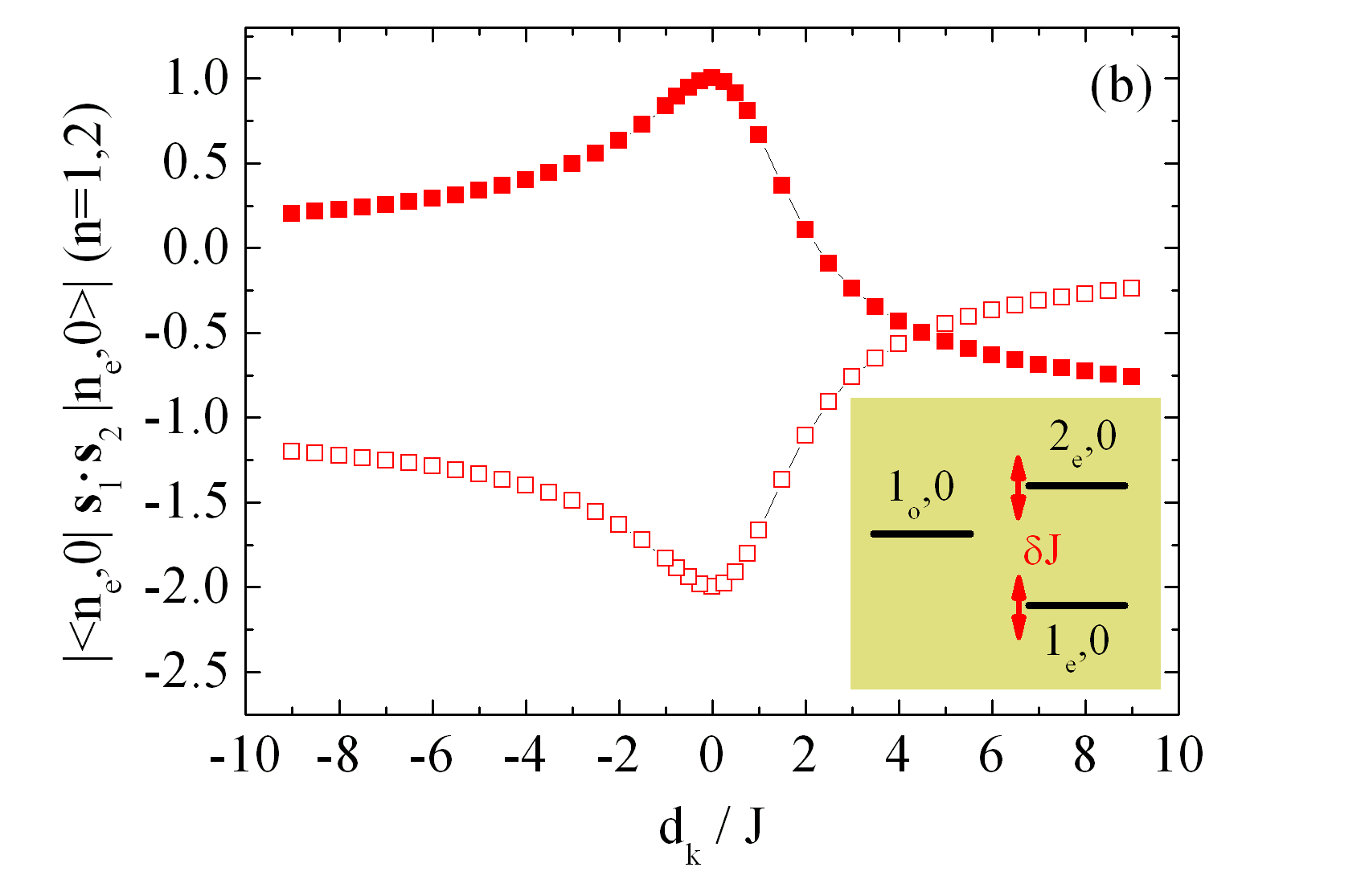}
\caption{\label{fig2} Diagonal matrix elements of the operators $s_{k,z}^2$ (a) and ${\bf s}_1 \cdot {\bf s}_2$ (b) in the spin dimer with $s_1=s_2=1$, as a function of the ratio between the anisotropy ($d$) and exchange ($J$) parameters. These matrix elements determine the renormalization of the energy differences between the eigenstates $|\beta_\xi,M\rangle$ (with $\xi = e,o$), induced by the modulation of $d$ or $J$ (figure insets). Empty and filled squares correspond to the ground ($n_e=1$) and second excited state ($n_e=2$), respectively.}
\end{figure}

The modulation of the $d$-factor can induce resonant transitions between the ground and the second excited states of the two-spin Hamiltonian, $|0,0\rangle$ and $|2,0\rangle$. The relevant matrix element is given by (see Appendix C)
\begin{equation}\label{eq08}
\langle 0,0|s^2_{k,z}|2,0\rangle = [s(s+1)(2s-1)(2s+3)/45]^{1/2},
\end{equation}
and thus scales quadratically with the length of the constituent spins. We remind that in this case, the two spins are not required to be inequivalent, {\it i.e.} the transition amplitude is finite also if $\delta d_1 = \delta d_2$. The renormalization of the $d$-factor can also modify the energy gap between the eigenstates. In particular, one can show that, while the expectation value of $s_{k,z}^2$ (with $k=1,2$) in the case of the singlet is, by symmetry, 
$
\langle 0,0|s^2_{k,z}|0,0\rangle = s(s+1)/3 ,
$
that for the triplet and the quintuplet state are given respectively by
\begin{equation}
\langle 1,0|s^2_{k,z}|1,0\rangle = (3s^2+3s-1)/5
\end{equation}
and 
\begin{equation}
\langle 2,0|s^2_{k,z}|2,0\rangle = (11s^2+11s-15)/21,
\end{equation}
as detailed in Appendix C.

\subsection{Spin dimer qubit with strong anisotropy}

If the anisotropy is not negligibly small with respect to the exchange interaction, $S$ is no longer a good quantum number. This implies, in general, that a renormalization of the exchange interaction can induce resonant transitions between the eigenstates of $H_a+H_b$, which we write as $|\beta_\xi,M\rangle$. Here, the index $\beta_\xi$ orders the system eigenstates within the subspace formed by states with total spin projection $M$ and even ($\xi = e$) or odd ($\xi = o$) values of $S$. In fact, as shown by Eq. \ref{eq04}, the operators $H_b$ doesn't couple states with even and odd values of the total spin. 

As far as the resonant transitions are concerned, such reduction of the system symmetry has three main implications. First, the selection rule $|\Delta S|=1$, which characterizes the transitions induced by the modulation of the $g$-factor [Eq. (\ref{eq02})], is replaced by the condition that the initial and final states must have different parities ({\it i.e.} different values of $\xi$). Second, the selection rules $\Delta S=0$ [Eq. (\ref{eq03})] and $|\Delta S|=0,2$ [Eq. (\ref{eq04})] are both replaced by the condition that the initial and final states must have same parity. Third, transitions between different eigenstates of the spin-dimer Hamiltonian can be induced by the modulation of the exchange interaction, within a subspace of given $\xi$. 

The latter point is further investigated numerically in the simplest case of interest, namely that where $s=1$. Here, the Hamiltonian eigenstates $|\beta_\xi,M\rangle$ with $M=0$ are $|1_e,0\rangle$ and $|2_e,0\rangle$, belonging to the subspace $S=0,2$, and $|1_o,0\rangle = (|1,-1\rangle - |-1,1\rangle) /\sqrt{2}$, belonging to the subspace $S=1$ (the states on the right-hand side of the equation are expressed in the basis $|m_1,m_2\rangle$). 

The transition amplitudes corresponding to $\delta H_{a,g}$,  $\delta H_{a,J}$ and $\delta H_{b,d}$ depend on the matrix elements of the operators $s_{k,z}$, ${\bf s}_1 \cdot {\bf s}_2$, and $s_{k,z}^2$, respectively. These are plotted in Fig. \ref{fig1} as a function of $d/J$ (where $d\equiv d_1=d_2$). 
In the weak-anisotropy limit ($|d|/J \ll 1$), one recovers the results given in Eqs. (\ref{eq09}-\ref{eq08}) and a vanishing transition amplitude associated to the exchange operator.
In the limit of a large, easy-axis anisotropy ($-d/J \gg 1$), the ground state is $|1_e,0\rangle \simeq (|1,-1\rangle + |-1,1\rangle) /\sqrt{2}$, and can be resonantly coupled to $|1_o,0\rangle$ and $|2_e,0\rangle$, respectively by the modulation of the $g$-factor and of the exchange coupling. In the limit of a large, hard-axis anisotropy ($d/J \gg 1$), the ground state is $|1_e,0\rangle \simeq |0,0\rangle$, and can be resonantly coupled to $|2_e,0\rangle$, again by modulating the exchange coupling. Altogether, the simultaneous modulation of the exchange coupling and of the axial anisotropy allows the realization of resonant transitions between the ground and the second excited states in all the range of values of $d/J$.

\begin{figure}
\centering
\includegraphics[width=0.85\columnwidth]{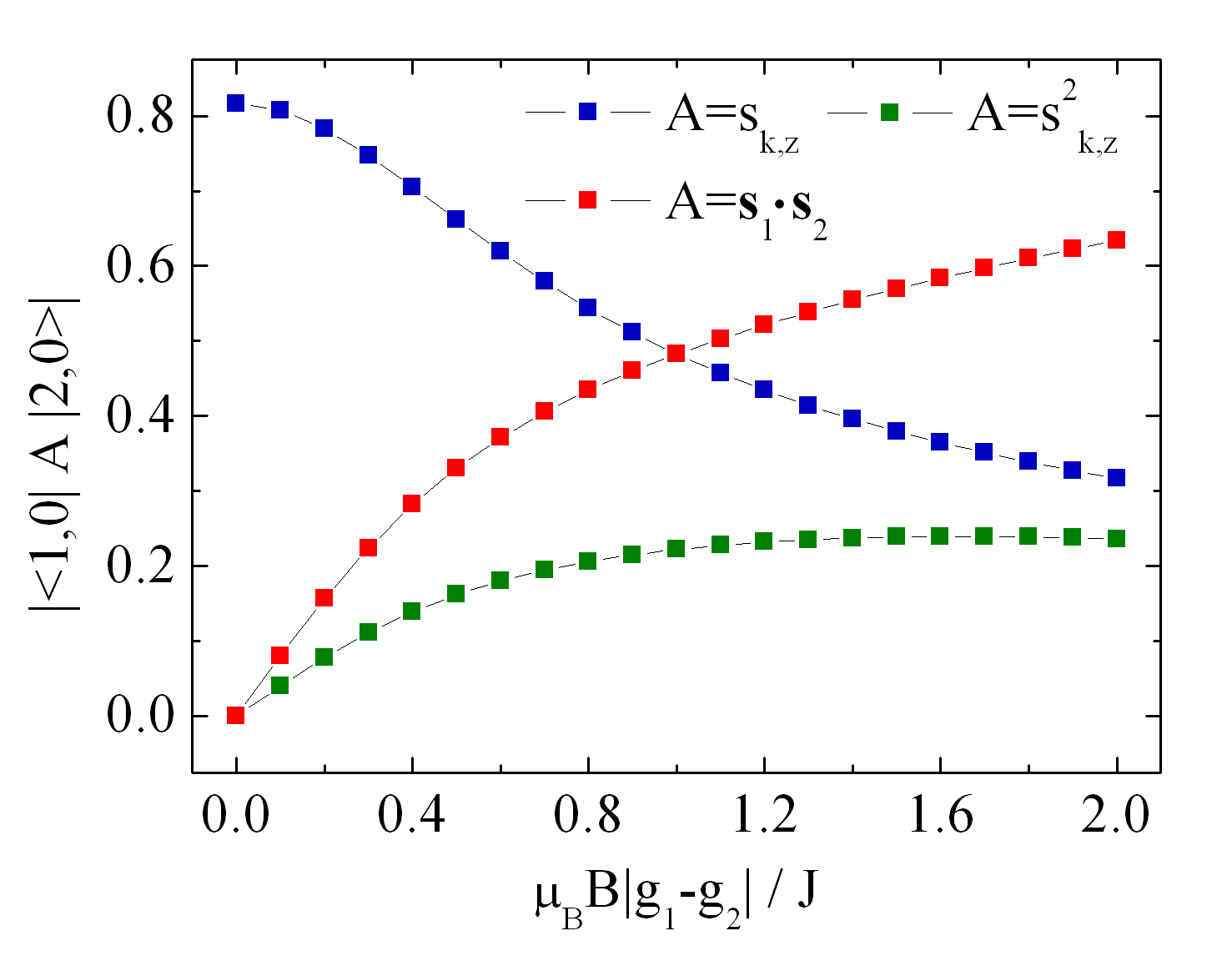}
\caption{\label{fig3} Matrix elements (in modulus) of the operators $s_{k,z}$ (blue), $s_{k,z}^2$ (green), and ${\bf s}_1 \cdot {\bf s}_2$ (red) between the two lowest eigenstates of the spin dimer with $s_1=s_2=1$, as a function of the ratio between the difference in the Zeeman energies of the two spins and the exchange coupling. The matrix elements determine the amplitudes of the transition between eigenstates $|\gamma,M\rangle$, induced by the modulation of $g$, $d$, or $J$.}
\end{figure}

An analogous analysis can be performed for the expectation values of the operators ${\bf s}_1 \cdot {\bf s}_2$ and $s_{k,z}^2$, which can be used in the nonresonant manipulation. Their values are plotted in Fig. \ref{fig2} as a function of the ratio $d/J$, for the eigenstates $|1_e,0\rangle$ and $|2_e,0\rangle$. The values for the third eigenstate $|1_o,0\rangle$ are independent on the value of the anisotropy, and are given respectively by
$\langle s_{k,z}^2 \rangle = 1$ and $\langle {\bf s}_1 \cdot {\bf s}_2 \rangle = -1$. The energy change $\delta E$ induced by changes both in the anisotropy and in the exchange interaction are all different for moderate values of $|d|/J$, such that the condition in Eq. (\ref{nonres}) is satisfied for all pairs of eigenstates. In the large-anisotropy limit, the energy renormalizations of the eigenstate $|1_o,0\rangle$ tend to coincide with that of either $|1_e,0\rangle$ or $|2_e,0\rangle$ (depending on the sign of $d$), thus making the gap between the two states insensitive to modulations of both $d$ and $J$. 

Overall, the above results show that, in the presence of $S$ mixing, the modulation of $d$ and $J$ allows one to implement both resonant and nonresonant manipulation between the two eigenstates of the spin dimer belonging to the even-$S$ subspace ($\xi =e$) at all values of $d/J$. Besides, the modulation of $g$ resonantly couples the ground and first excited states for $-d/J \gtrsim 1$.

\subsection{Spin dimer qubit with strongly inhomogeneous $g$-factors}

A certain degree of $S$-mixing in the eigenstates of the spin dimer is also induced by inhomogeneous values of (couplings to) the magnetic field. The resulting differences between the Zeeman energies of the two spins can be exploited in the manipulation of the singlet-triplet qubit \cite{Levy2002a,Petta2005a,Maune2012a}. In fact, for $g_1 \neq g_2$ the Hamiltonian $ H = J {\bf s}_1 \cdot {\bf s}_2 + \mu_B B \sum_{k=1}^2 g_k s_{k,z} $ commutes with $S_z$, but not with ${\bf S}^2$. Moreover, within the $M=0$ subspace, one can also write: $ H = J {\bf s}_1 \cdot {\bf s}_2 + \frac{1}{2}\mu_B B (g_1-g_2) (s_{1,z}-s_{2,z}) $, because the remaining part of the Zeeman Hamiltonian vanishes identically. Therefore, the eigenstates can be written in the form $|\gamma,M\rangle$, with $\gamma=1,2,3$, and only depend on the ratio between $\mu_B B (g_1-g_2)$ and $J$.

For simplicity, we only discuss the transition between the ground and the first excited state, whose amplitude we plot in Fig. \ref{fig3} as a function $\mu_B B |g_1-g_2|/J$. The results show that a significant transition amplitude can be obtained by the modulation of the $g$-factors (blue). We note that, as in the previously considered cases, such modulation needs to be different for the two spins ($\delta g_1 \neq \delta g_2$) in order for the transition amplitude to be finite, because also in this case $H$ commutes with $S_z$. The transition amplitudes related to the axial anisotropy (green) and to the exchange interaction (red), which vanish for $g_1=g_2$, increase monotonically with increasing $|g_1-g_2|$ in the displayed range of values (whereas they decrease and eventually tend to zero in the limit $\mu_B B |g_1-g_2| \gg J$, not shown).

Therefore, the modulation of the exchange interaction can be used to induce resonant transitions also in the spin dimer, if the eigenstates exhibit a significantly degree of $S$-mixing, due to the difference between the Zeeman energies of the two spins.

\subsection{Exchange-only approach in a spin square qubit}

If $S$ is a good quantum number, the resonant spin manipulation through exchange interaction requires a spin cluster larger than the dimer. One of the simplest systems we can possibly consider is a square ($N=4$) of spins $s=1$, with identical exchange couplings $J$ between nearest neighbors. In fact, a triangle of $s$ spins always presents only one eigenstate with $S=0$. In a square of $s=1/2$ spins, the only excited singlet state is degenerate with two triplet states, and is thus unsuitable for forming, together with the ground state, an effective two-level system. 

In the case of the square of $s=1$ spins, any exchange operator involving neighboring spins couples the two lowest eigenstates $|\alpha,S,M\rangle$ belonging to the $S=0$ subspace. In fact, by using the analytical expressions of these eigenstates (see Appendix D), one can show that 
\begin{equation}
\langle 1,0,0 | {\bf s}_k \cdot {\bf s}_{k+1} | 2,0,0\rangle| = (-1)^{k+1}\sqrt{5}/(2\sqrt{3}),
\end{equation}
where ${\bf s}_{N+1} \equiv {\bf s}_1$. Combining the above equation with Eq. (\ref{eq11}), one obtains the following expression for the transition amplitude:
\begin{equation}
\langle 0 |\delta H_{a,J} (t)| 1 \rangle 
= \frac{\sqrt{5}}{4\sqrt{3}} \sum_{k=1}^4 (-1)^{k+1} \delta J_{k\,k+1}(t) ,
\end{equation}
where $J_{45}\equiv J_{41}$. This implies that the effect of the applied field on the exchange couplings should not be too symmetric. If, for example, the renormalizations $\delta J_{k\, k+1}$ are all identical, their contributions cancel out and the transition amplitude vanishes.

The two eigenstates also present different expectation values of the exchange operator, as required for using the renormalization of the exchange couplings for the nonresonant manipulation:
\begin{equation}
\langle 1,0,0 | {\bf s}_k \cdot {\bf s}_{k+1} | 1,0,0\rangle \!=\! 
3\langle 2,0,0 | {\bf s}_k \cdot {\bf s}_{k+1} | 2,0,0\rangle \!=\! -3/2.
\end{equation}

\section{Discussion}

In the present Section we discuss analogies and differences with respect to analogous approaches, and the main open issues related to the spin-electric manipulation of molecular spin clusters. 

\subsection{Differences with respect to the single spin case}

The modulation of the $g$-factor, of the axial anisotropy, and of the exchange interaction has been investigated experimentally and proposed as a suitable means for coherently manipulating the spin states in molecular and solid-state systems. 

In the $g$-tensor modulation resonance, an oscillating electric field modulates the coupling of the spin to a static magnetic field, emulating the effect of an oscillating magnetic field and inducing resonant transitions between different Zeeman levels \cite{Crippa2018a,Pingenot2008a,Kato2003a,Takahashi2013a,Schroer2011a}. This differs from what we discuss in the present paper both in terms of the spin manipulation that one aims at and of the properties that the system should possess in order to make such manipulation possible. In $g$-tensor modulation resonance one induces, by different means, the same kind of transitions as in electron spin resonance, namely those between states with different spin projection. Here, instead, the modulation of the $g$-factor induces transitions between states with identical spin projection and different values of the total spin or of some other scalar quantity (such as the partial spin sums or the spin chirality) that commutes with $S_z$. Quantities such as the total or the partial spin sum can be modified or defined within a spin cluster, but not in individual spins, where the present approach cannot be implemented. In both cases, in order to induce resonant transitions one needs that the static and the time-dependent Hamiltonians don't commute. In $g$-tensor modulation resonance, this requirement is met if the field-induced variation of the $g$-tensor is anisotropic, whereas here the modulation of the $g$-factors must be inhomogeneous throughout the spin cluster. 

Analogously, an electric-field induced modulation of the axial anisotropy has been demonstrated and exploited for implementing both a nonresonant and a resonant manipulation of individual spins \cite{George2013a}. Here we show how this kind of spin-electric coupling can be used to manipulate quantum numbers related to collective properties of a spin-cluster state.

\subsection{Other cases of exchange modulation}

The modulation of the exchange interaction has been proposed as a general approach for the implementation of quantum computing in the $M=0$ subspace. To this aim, one needs to encode each logical qubit in the state of more than one physical qubit and achieve a high degree of control on each exchange coupling \cite{DiVincenzo2000a,Levy2002a}. The singlet-triplet qubits in coupled semiconductor quantum dots essentially follow this approach \cite{Petta2005a,Maune2012a}. In the present paper, we consider the spin manipulation through the modulation of the exchange interaction in a related and yet different perspective. In fact, in molecular spin clusters one cannot expect to control each exchange coupling individually, and the relative changes induced by the applied electric fields can be rather small \cite{Islam2010a,Nossa2012a}. Besides, the magnetic environment (and especially the nuclear spins) couple locally, and not collectively, to the spins that form the cluster. In order to protect the state from decoherence, one thus needs to fulfill more stringent conditions than that concerning the total spin projection \cite{Troiani2012a}. Given these premises, we aim here at identifying pairs of eigenstates of the spin cluster that one can couple resonantly and nonresonantly by pulsed electric fields, thus implementing orthogonal rotations within an effective two-level system. We also note that the spin Hamiltonians of molecular nanomagnets generally contain different terms. Therefore, the desired rotations need not be implemented through the modulation of the exchange interaction alone, but can result from the combined renormalization of different parameters. 

The modulation of the exchange interaction has also been proposed for the manipulation of spin chirality qubits, and specifically in odd-numbered rings of half-integer spins with antisymmetric exchange interaction \cite{Trif2008a,Trif2010a}. The present results show that this kind of spin-electric coupling can in fact be used to achieve for electrically manipulating the spin in decoherence-free subspaces within a wider class of systems, characterized by significantly different geometries and spin Hamiltonians. We also note that the present approach doesn't necessarily require the use of a static magnetic field. In fact, here this is only required in order to remove the degeneracy between the $M=0$ and the $|M|>0$ states within the $S=1$ and $S=2$ subspaces, in those cases where such multiplets are involved. However, such degeneracy can also be removed by the axial anisotropy, which would make the magnetic field  in principle unnecessary.

\subsection{Experimental detection of the spin-electric couplings}

At least two major aspects need to be further investigated. The first one has to do with the detection of transitions that result in a change of the total spin. This tends to be less straightforward than the detection of a change in the magnetization, such as those induced in electron spin resonance. One possible solution consists in coupling the ensembles of spin clusters to  superconducting resonators \cite{Xiang2013a}. The strong coupling of such a resonator with ensembles of molecular spins has already been demonstrated by using the magnetic component of the field \cite{Bonizzoni2016a,Bonizzoni2017a}. Analogous experiments involving the electric component of the field can be performed to demonstrate the spin-electric coupling, mediated by the modulation of any parameter in the spin Hamiltonian \cite{George2013a}. Electron spin resonance in the presence of an applied electric field also represents a useful tool for investigating the spin-electric coupling in different classes of systems \cite{Boudalis2018a,Fittipaldi2019a,Liu2019a}. In this respect, one should however consider that this approach gives access to transitions and linear superpositions that differ from the ones we focus on here, and that a given spin-electric coupling can produce an effect in one case and not in the other. The second aspect has to do with the identification of systems that are amenable to electric manipulation, and with the estimate of the related coupling constants. In this respect, indications for a rational design of the system can come from ab initio calculations that explicitly take into account the orbital degrees of freedom \cite{Baadji2009a,Islam2010a,Nossa2012a,Bellini2008a,Chiesa2013a}, and by the related field of multiferroics \cite{Khomskii2009a,Matsukura2015a}.

\subsection{Decoherence mechanisms}

The fact that the relevant eigenstates of the molecular spin cluster are characterized by a vanishing expectation value of the single- and total-spin operators implies that the qubit system is substantially decoupled from the magnetic noise, mainly resulting from the interaction with nuclear spins and with the electron spins of neighboring molecules. On the other hand, the spin-electric coupling introduces novel decoherence channels, related to electric noise. 

Electric noise is ubiquitous in solid-state devices, and can have different physical origins \cite{Connors2019a,Langsjoen2012a,Paladino2014a}. For example, dynamical charge traps generally induce low-frequency noise, which might affect the spin-cluster qubits mainly by inducing fluctuations in the energy gaps, and thus dephasing. The energy fluctuations induced by the electric noise are expected to be smaller than those typically induced by the magnetic environment, because the spin-electric coupling is expected to be smaller than the corresponding coupling $g\mu_B$ between the spin and the magnetic field. In fact, degrees of freedom that are weakly coupled to the control knobs (and thus not amenable to fast manipulation) typically tend to be weakly coupled to the environment (and thus coherent). At this stage, it's difficult to draw conclusions on such coherence-controllability trade-off in molecular spin clusters. As mentioned above, further investigation is required in order to understand what values of the different spin-electric couplings, and thus of the renormalizations of $J$, $g$, or $d$, can actually be achieved. Also, intrinsic sources of electric noise - typically represented by localized charge states that fluctuate between two (or more) configurations - need to be characterized, in order to optimize the operations and improve the coherence, as done for singlet-triplet qubits in semiconductor quantum dots \cite{Dial2013a}.

\section{Conclusions}

In conclusion, we have investigated the state manipulation in spin clusters by means of electric fields. In particular, we have considered the manipulation of the total spin $S$ and of other scalar quantities within the $\langle s_{k,z}\rangle=0$ (and thus also $M=0$) subspace, in view of the potential advantages that this might present in terms of decoherence. We have shown that the modulation of different parameters entering the spin Hamiltonian ($g$-factor, axial anisotropy, exchange coupling) can be used for implementing both resonant and a nonresonant manipulation schemes, and have derived the relevant selection rules. Finally, we have shown that $g$- and $d$-modulation induced transitions can in principle take place in a system as simple as a spin dimer. Transitions induced by the modulation of the exchange interaction can also be observed in the dimer in the presence of $S$-mixing, or otherwise in more complex spin clusters (such as a spin square).

The author acknowledges financial support from the European Commission through the project {\it IQubits} (Call: H2020-FETOPEN-2018-2019-2020-01, Funding scheme: Research and Innovation Action, Grant agreement ID: 829005).


\section*{APPENDIX A: SELECTION RULES RESULTING FROM THE WIGNER-ECKART THEOREM}\label{appA}

From the Wigner-Eckart theorem it follows that the matrix element of a spherical tensor operator $\langle S_0, M_0 | T_q^{(p)} | S_1,M_1\rangle $ is proportional to the Clebsch-Gordan coefficient $\langle S_1\, M_1\, p\, q |S_0\, M_0\rangle$, where $p$ is the rank of the spherical tensor and $q$ specifies the component \cite{Gatteschi2006a}. Therefore, it can be different from zero only if the spin projections fulfill the equation
\begin{equation}\label{a2}
M_0+q-M_1=0
\end{equation} 
and the values of the total spin satisfy the inequalities
\begin{equation}\label{a1}
|S_0-p|\le S_1\le S_0+p.
\end{equation} 

The selection rules reported in this paper are derived from the above relations, applied to the cases of the operators $s_{k,z}$, $s_{k,z}^2$, and ${\bf s}_k \cdot {\bf s}_l$ (with $k,l = 1,\dots,N$). All these operators can be expressed in terms of $q=0$ components of spherical tensor operators. This implies, for the total spin projections, that $M_0=M_1$. 

The single-spin projection operators $s_{k,z}$ represent the $0$-th component of a rank $p=1$ spherical tensor operator. According to the above Eq. \ref{a1}, this implies that $|\Delta S|= |S_1-S_0| \le 1$ for the transitions induced by the modulation of the $g$-factor. 

The squared single-spin projections $s_{k,z}^2$ can be written as the sum of two spherical tensor operators, with ranks $p=0$ and $p=2$. The former one can be identified with the constant $s_k(s_k+1)/3$; the latter one thus corresponds to $s_{k,z}^2-s_k(s_k+1)/3$. Equation \ref{a1} thus implies that that $|\Delta S| \le 2$ for the transitions induced by the modulation of the axial anisotropy.

Finally, the exchange terms ${\bf s}_k \cdot {\bf s}_l$ are spherical tensor operators of rank $p=0$. Therefore, Eq. \ref{a1} translates in the selection rule $\Delta S =0$ for the transitions induced by the modulation of the exchange couplings.

\section*{APPENDIX B: SELECTION RULES RESULTING FROM SYMMETRY ARGUMENTS}\label{appB}

In the present paper, we focus on transitions between states $|\alpha,S,M\rangle$ with arbitrary $S$ and $M=0$. These states are also eigenstates of the unitary operators corresponding to $\pi$ rotations around any axis that lies in the $xy$ plane. In fact, one can show that 
\begin{equation}
e^{-iS_x\pi/\hbar} | \alpha,S,M=0\rangle = (-1)^S | \alpha,S,M=0\rangle .
\end{equation}
The powers of the single-spin projection operators have defined symmetries with respect to such rotations: 
\begin{equation} 
e^{iS_x\pi/\hbar} s_{k,z}^r e^{-iS_x\pi/\hbar} = (-1)^r s^r_{k,z}.
\end{equation} 
As a result, by applying the rotations either to the operator or to the states in the expression of the matrix element, one obtains that:
\begin{eqnarray}
(-1)^{S_0+S_1} \langle S_0 | s^r_{k,z} |S_1\rangle 
=
(-1)^{r} \langle S_0 | s^r_{k,z} |S_1\rangle ,
\end{eqnarray}
where we have omitted the quantum numbers $\alpha$ and $M$, in order to simplify the expressions. From the equation above, it follows that the matrix element of $s_{k,z}^r$ always vanishes if $r$ is odd and $S_0+S_1$ is even and if $r$ is even and $S_0+S_1$ is odd. 

\section*{APPENDIX C: EIGENSTATES OF THE SPIN DIMER}\label{appC}

The singlet state of a two-spin system, expanded on the basis $\{ |m_1,m_2\rangle \}$, is given by the following expression:
\begin{equation}
|0,0\rangle = \frac{1}{\sqrt{C_0}} \sum_{m=-s}^s (-1)^{s+m} |m,-m\rangle ,
\end{equation}
where
$s$ is the length of each of the two spins and $C_0^{2} = 2s+1$. In view of the selection rules discussed above,
the triplet state with $M=0$ can be easily obtained by applying the $s_{k,z}$ operator (with $k=1,2$) to the above singlet, and then normalizing. The result is given by:
\begin{equation}
|1,0\rangle = \frac{1}{\sqrt{C_1}} \sum_{m=-s}^s (-1)^{s+m} m |m,-m\rangle ,
\end{equation}
where 
$C_1^{2} = s(s+1)(2s+1)/3 $.
Along the same lines, the state with $S=2$ and $M=0$ can be obtained by applying to the singlet the component of $s_{z,k}^2$ that corresponds to a spherical tensor operator of rank 2, {\it i.e.} $s_{z,k}^2-s(s+1)/3$, and then normalizing. The result reads:
\begin{equation}
|2,0\rangle = \frac{1}{\sqrt{C_2}} \sum_{m=-s}^s (-1)^{s+m} [m^2-s(s+1)/3] |m,-m\rangle ,
\end{equation}
where 
$C_2^{2} = s(s+1)(2s+1)(2s-1)(2s+3)/45 $.

Given the above expressions of the relevant eigenstates, one can derive the analytical expressions of the diagonal and off-diagonal matrix elements that are given in Sect. V. In the derivation, we make use of the equations 
$\sum_m m^2=s(s+1)(2s+1)/3$,
$\sum_{m=-s}^s m^4 = s(s+1)(2s+1)(3s^2+3s-1)/15 $, 
and 
$\sum_{m=-s}^s m^6 = s(s+1)(2s+1)(3s^4+6s^3-3s+1)/21 $, 
which result from the Faulhaber's formulas.

\section*{APPENDIX D: EIGENSTATES OF THE SPIN SQUARE}\label{appD}

The singlet ground state of the square of $s=1$ spins with identical exchange coupling $J$ between nearest neighbors is given by:
\begin{equation}
|1,0,0\rangle = \frac{1}{6\sqrt{5}} \sum_{p=1}^5 \alpha_p \sum_k^{(p)} |m_{1,kp},m_{2,kp},m_{3,kp},m_{4,kp}\rangle .
\end{equation}
The states $|m_1,m_2,m_3,m_4\rangle$ belonging to the five groups ($p=1,\dots 5$) are characterized as follows: 
$(i)$ two neighboring spins with projection $1$, the other two with projection $-1$ ($\alpha_1=1$); 
$(ii)$ two non-neighboring spins with projection $0$, the other two with projection $+1$ and $-1$ ($\alpha_2=2$);
$(iii)$ two neighboring spins with projection $0$, the other two with projection $+1$ and $-1$ ($\alpha_3=3$); 
$(iv)$ all spins with projection $0$ ($\alpha_4=4$);
$(v)$ two classical ground states of the antiferromagnetic spin ring, with antiparallel nearest neighbors ($\alpha_5=6$). 

The first excited state belonging to the $S=0$ subspace is given by:
\begin{equation}
|2,0,0\rangle = \frac{1}{2\sqrt{3}} \sum_{p=1}^4 \beta_p \sum_k^{(p)} |m_{1,kp},m_{2,kp},m_{3,kp},m_{4,kp}\rangle .
\end{equation}
The states $|m_1,m_2,m_3,m_4\rangle$ belonging to the four groups ($p=1,\dots 4$) are characterized as follows: 
$(i)$ spins 1 and 2 (3 and 4) with projection $+1$, spins 3 and 4 (1 and 2) with projection $-1$ ($\beta_1=1$); 
$(ii)$ spins 2 and 3 (1 and 4) with projection $+1$, spins 1 and 4 (2 and 3) with projection $-1$ ($\beta_2=-1$); 
$(iii)$ spins 1 and 2 (3 and 4) with projections $\pm 1$, spins 3 and 4 (1 and 2) with projection $0$ ($\beta_3=1$); 
$(iv)$ spins 2 and 3 (1 and 4) with projections $\pm 1$, spins 1 and 4 (2 and 3) with projection $0$ ($\beta_4=-1$).


\end{document}